\documentstyle[aps]{revtex}
\begin{document}
\draft
\title{Exact Eigenstates and Magnetic Response of Spin-1 and Spin-2
Vectorial Bose-Einstein Condensates}
\author{Masato Koashi$^1$ and Masahito Ueda$^2$}
\address{$^1$NTT Basic Research Laboratories, 3-1 Morinosato Wakamiya,
Atsugi, Kanagawa 243-0198, Japan\\
$^2$Department of Physical Electronics, Hiroshima University,
Higashi-Hiroshima 739-8527, Japan}
\maketitle
\begin{abstract}
The exact eigenspectra and eigenstates of spin-1 and spin-2 vectorial
Bose-Einstein condensates (BECs) are found, and their response to   
a weak magnetic field is studied and compared with their mean-field    
counterparts.
Whereas mean-field theory predicts the vanishing population of the  
zero magnetic-quantum-number component of a spin-1 antiferromagnetic 
BEC, the component is found to become populated as the magnetic field    
decreases. The spin-2 BEC exhibits an even richer magnetic response  
due to quantum correlation between 3 bosons.
\end{abstract}
\pacs{PACS numbers: 03.75.Fi, 05.30.Jp}


Bose-Einstein condensates (BECs) of alkali-metal atoms have internal 
degrees of freedom due to the hyperfine spin of the atoms. These     
degrees of freedom are frozen in a magnetic trap, but an optical trap
liberates them to allow BEC to be in a superposition of magnetic     
sublevels~\cite{Stamper-Kurn}.  BEC is therefore described by a     
vectorial rather than scalar order parameter. A new feature in this  
BEC system as compared to superfluid helium-three is the fact that   
its response to an external magnetic field is dominated by electronic   
rather than nuclear spin, and hence the response is much stronger    
than that of superfluid helium-three.
This opens up possibilities of manipulating the magnetism of superfluid  
vapors. Observation of spin domains by an MIT group~\cite{Stenger}   
offers an excellent example of such manipulations.                 
While the experiments reported so far achieved only the spin-1
vectorial BEC, the spin-2 BEC also appears feasible by using
the $F=2$ multiplet of bosons 
such as $^{23}$Na, $^{87}$Rb, or $^{85}$Rb.

The mean-field theory (MFT) for describing a vectorial BEC was developed
independently by Ohmi and Machida~\cite{Ohmi} and by Ho~\cite{Ho} by 
generalizing the Gross-Pitaevskii equation under the restriction of  
gauge and spin-rotation symmetry; they also used it to predict       
various spin textures and topological excitations.
Law {\it et al.}~\cite{Law} utilized techniques developed in quantum 
optics~\cite{Eberly,Wu} to study many-body states of spin-1 BEC in   
the absence of external fields, and found that spin-exchange         
collisions lead to rather complicated dynamical behavior of BEC   
that MFT fails to capture.  
In this Letter, we study magnetic response of spin-1 and  spin-2 BECs
by explicitly constructing exact eigenspectra and eigenstates, and   
compare the results with their mean-field counterparts.

We first consider a system of spin-1 bosons interacting via s-wave   
scattering. The second-quantized  Hamiltonian of the bosons subject  
to a uniform magnetic field $\bbox{B}$ and in a confining potential  
$U(\bbox{r})$ is given by
\begin{eqnarray}
\hat{H}_0=\!\int \!d\bbox{r}\left[
\frac{\hbar^2}{2M}\bbox{\nabla}\hat{\Psi}_\alpha^\dagger
\cdot\bbox{\nabla}\hat{\Psi}_\alpha
+U\hat{\Psi}_\alpha^\dagger\hat{\Psi}_\alpha
+\frac{\bar{c}_0}{2} \hat{\Psi}_\alpha^\dagger\hat{\Psi}_\beta^\dagger
\hat{\Psi}_\beta\hat{\Psi}_\alpha
\right.
\nonumber \\
\left.  +\frac{\bar{c}_1}{2} \hat{\Psi}_\alpha^\dagger
\hat{\Psi}_\beta^\dagger
\bbox{f}_{\alpha\alpha^\prime}
\cdot\bbox{f}_{\beta\beta^\prime}
\hat{\Psi}_{\beta^\prime}\hat{\Psi}_{\alpha^\prime}
\!-\!g\mu_B\hat{\Psi}_\alpha^\dagger
\bbox{B}\cdot\bbox{f}_{\alpha\alpha^\prime}
\hat{\Psi}_{\alpha^\prime}
\right],
\label{h0}
\end{eqnarray}
where $M$ is the mass of the bosons,
$\hat{\Psi}_\alpha$ describes their
field  operator
with magnetic quantum number $\alpha=-1,0,1$,
and $\bar{c}_0$ and $\bar{c}_1$ are related to
scattering lengths $a_0$ and $a_2$ of two colliding
bosons with total angular momentum $0$ and $2$ by
$\bar{c}_0=4\pi\hbar^2(2 a_2+a_0)/3M$ and
$\bar{c}_1=4\pi\hbar^2(a_2-a_0)/3M$~\cite{Ho}.
Here and in the rest of this Letter,
it is assumed that
repeated indices are to be summed, and that
the total number $N$ of bosons
in the system is fixed.
We further assume that the
external
magnetic field is weak and $|c_1|\ll c_0$
so that
the coordinate wave function $\phi({\bbox r})$ is
independent of the state of the spin space and
solely
determined by the
first three terms of Eq.\ (\ref{h0}), namely,
\begin{equation}
\left[
-\frac{\hbar^2\bbox{\nabla}^2}{2M}+U
+\bar{c}_0(N-1)|\phi|^2\right]\phi
=\epsilon\phi.
\end{equation}
Substituting $\hat{\Psi}_\alpha=\hat{a}_\alpha\phi$ into Eq.\ (\ref{h0})
and keeping only spin-dependent terms, we obtain
\begin{equation}
\hat{H}=
\frac{c_1}{2}:\hat{\bbox{F}}\cdot\hat{\bbox{F}}:-p\hat{F}_z,
\label{h1}
\end{equation}
where $c_1\equiv \bar{c}_1\int d\bbox{r}|\phi|^4$,
$:\hat{O}:$
arranges the operator $\hat{O}$ in normal order, and the three 
components $\hat{F}_{x,y,z}$ of the hyperfine-spin operator     
$\hat{\bbox{F}}$ are defined in terms of $3\times3$ spin-1
matrices $F_{x,y,z}$ as 
$\hat{F}_x=(F_x)_{\alpha\beta}\hat{a}_\alpha^\dagger\hat{a}_\beta$, 
etc. In the following discussions we assume that $p\equiv g\mu_B B>0$.

Exact energy eigenstates and eigenvalues of Hamiltonian (\ref{h1}) 
can be constructed as follows. We introduce an operator             
$\hat{A}^\dagger
 \equiv[(\hat{a}^\dagger_0)^2-2\hat{a}^\dagger_1\hat{a}^\dagger_{-1}]/\sqrt{3}$
which creates a pair of bosons in the spin-singlet state when operated
on the vacuum, and define a set of states $|N_2,F,F_z\rangle$ as
\begin{equation}
|N_2,F,F_z\rangle\equiv
Z^{-1/2}(\hat{A}^\dagger)^{N_2}(\hat{F}_-)^{F-F_z}
(\hat{a}_1^\dagger)^F|vac\rangle,
\end{equation}
where  $Z$ is the normalization constant and $\hat{F}_-$ is the 
lowering operator for $F_z$. 
Since $\hat{A}^\dagger$ commutes with $\hat{\bbox{F}}^2$
and
$\hat{F}_z$, $|N_2,F,F_z\rangle$ is the simultaneous eigenstate of
$\hat{N}$, $\hat{\bbox{F}}^2$,
and $\hat{F}_z$, with total number of bosons $N=F+2N_2$, total
spin $F$, and magnetic quantum number $F_z$. This state is
thus an energy eigenstate of $\hat{H}$ with energy eigenvalue
\begin{equation}
E=\frac{c_1}{2}[F(F+1)-2N]-pF_z.
\label{E1}
\end{equation}
The number of possible states $|N_2,F,F_z\rangle$
for a fixed $N$ is obtained as the coefficient of
$x^N$
of the generating
function
$
\sum_{N_2,F,F_z}x^{N}=\sum_{N_2,F}(2F+1)x^{2N_2+F}=(1-x)^{-3},
$
and is given by $(N+1)(N+2)/2$.
Since this number coincides with that of
linearly independent states for a system of
N spin-1 bosons, i.e., $_{N+2}C_2$, the set
$\{|N_2,F,F_z\rangle\}$ forms a complete orthonormal basis.

The ground state is obtained by minimizing Eq.(\ref{E1})
with $N$ held fixed.
When $c_1<0$, it is a ferromagnetic
state in which all bosons occupy the $m=1$ state, in agreement
with the prediction of MFT.
When $c_1>0$,
$|N_2=(N-F)/2,F,F_z=F\rangle$ is the exact ground state for
\begin{equation}
F-\frac{1}{2}<\frac{p}{c_1}
<F+\frac{3}{2}.
\label{step1}
\end{equation}
That is, magnetization increases stepwise,
 taking the
values
$F=N-2N_2$ with the step size $\Delta F=2$ as the magnetic field
increases. In contrast, according to MFT,
magnetization increases linearly with the magnetic
field as $F_z\sim [g\mu_B/c_1]B$. Both theories, however, predict
the same average slope.
The difference between the exact ground state energy E and
the minimum energy $E_M$ in MFT is
$E-E_M\sim -c_1(N-F)(N+F)/2N$,
which is of the order of the antiferromagnetic interaction energy between
one particular particle and the rest.

From the form of the ground states
$|N_2,F=F_z,F_z\rangle\propto(\hat{A}^\dagger)^{N_2}
(\hat{a}_1^\dagger)^F|vac\rangle$, we may say
that increasing the
magnetic field breaks singlet `pairs' one by one,
which results in
the stepwise increase of magnetization.
These `pairs' are in some
sense analogous to Cooper pairs of electrons or $^3$He,
but there is a remarkable difference. In the case of Cooper
pairs, the state is symmetric only under the permutations that
do not break any pairs. On the other hand,
in the present case the state is symmetric for any permutations of
constituent particles.

An observable that makes a striking distinction from MFT
is the population $n_0$ of the $m=0$ Zeeman sublevel, which is
predicted to be zero in MFT.
For the exact ground state, the expectation of $n_0$ can be calculated  as
\begin{equation}
\bar{n}_0\equiv\langle \hat{a}_0^\dagger\hat{a}_0\rangle
=\frac{2N_2}{2F+3}=\frac{N-F}{2F+3}.
\label{m0p}
\end{equation}
The nonzero value of $\bar{n}_0$ makes a sharp contrast
to the  prediction of MFT.
When $1\ll F \ll N$, $\bar{n}_0$ is inversely proportional
to the magnetic field as
$\bar{n}_0 \sim \bar{c}_1 \rho / (2g\mu_BB)$, where
$\rho\equiv N\int d\bbox{r}|\phi|^4$ is the average number density
of BEC. Note that when $B\neq 0$,
$\bar{n}_0/N$ is finite only in the mesoscopic regime, and
vanishes in the limit $N\rightarrow\infty$.
For sodium atoms in the $F=1$ state, where
$\bar{c}_1/\mu_B\sim 10^{15}$cm$^3$G, experiments of 
Ref.~\cite{Stamper-Kurn} achieved
$\rho\sim 10^{-19}$cm$^{-3}$ in an optical dipole trap.
From these values, we see for example, that in order to observe 
$\bar{n}_0$  of the order of $10^3$, a small
magnetic field of the order of $10^{-7}$G is required.

The rapid decrease of $\bar{n}_0$ as a function of $F$ can be ascribed
to
the indistinguishability of bosons.
If all particles were distinguishable,
the state could be written as
$
\Psi=\prod_{i=1}^{N_2}\Psi_{2i-1,2i}
\prod_{j=1}^{N-2N_2}|1\rangle_{2N_2+j},
$
where $\Psi_{i,j}$ is the spin-singlet state
for 
particles $i$ and $j$.
There are $N_2$ singlet pairs
so that the $m=0$ population would be $(N-F)/3$,
which decreases only linearly as $F$ increases.
The wavefunction of a bose system is obtained by the symmetrization
of $\Psi$. 
Adding bosons in the $m=1$ state increases the relative probability 
amplitudes having large $m=1$ occupation numbers. This is nothing but
the bosonic enhancement and may be interpreted as a consequence of  
the constructive interference among the permuted terms. 
The expectation value of the $m=0$ population thus decreases rapidly
towards the MFT value of zero
with the increasing magnetic field, as
can be seen from Eq.~(\ref{m0p}).

We next consider BEC of spin-2 bosons. Bose symmetry requires that
the total angular momentum of two colliding bosons is restricted to 0,
2, and 4, so that the interaction Hamiltonian which describes
binary collisions via the s-wave scattering is generally written as
$\hat{V}=g_4 \hat{P}_4+g_2 \hat{P}_2+g_0 \hat{P}_0$,
where $\hat{P}_F (F=0,2,4)$ denotes the projection operator for the
total angular momentum $F$~\cite{Ho}, $g_F$ is related to
scattering length $a_F$ by $g_F=4\pi\hbar^2a_F/M$,
and we have omitted the
coordinate delta function describing the contactness of the interaction.
Using $\hat{P}_4+\hat{P}_2+\hat{P}_0=\hat{1}$ and
$\hat{\bbox{f}}_i\cdot
\hat{\bbox{f}}_j=4\hat{P}_4-3\hat{P}_2-6\hat{P}_0$
($i$ and $j$
label particles), $\hat{V}$ is
rewritten as
$\hat{V}=\bar{c}_0+{\bar{c}_1\hat{\bbox{f}}_i\cdot\hat{\bbox{f}}_j}
+\bar{c}_2\hat{P}_0$,
where $\bar{c}_0=(3 g_4+4 g_2)/7$, $\bar{c}_1=(g_4- g_2)/7$
and $\bar{c}_2=(3 g_4- 10 g_2+7 g_0)/7$.

To derive the second-quantized form of the Hamiltonian,
it is convenient to introduce a new operator
$\hat{\cal S}_+\equiv
(\hat{a}^\dagger_0)^2/2
-\hat{a}^\dagger_1\hat{a}^\dagger_{-1}
+\hat{a}^\dagger_2\hat{a}^\dagger_{-2}$.
This operator creates, if applied to the vacuum,
a pair of bosons in the spin-singlet state. This pair, however,
should not be regarded as a single composite boson
because $\hat{\cal S}_+$ does not satisfy the commutation relations
for bosons. The operator $\hat{\cal S}_+$ instead
satisfies the $SU(1,1)$ commutation relations if we define
$\hat{\cal S}_-\equiv\hat{\cal S}_+^\dagger$ and
$\hat{\cal S}_z\equiv (2\hat{N}+5)/4$, namely,
$[\hat{\cal S}_z,\hat{\cal S}_\pm]=\pm\hat{\cal S}_\pm$
and
$[\hat{\cal S}_+,\hat{\cal S}_-]=-2\hat{\cal S}_z$;
the minus sign in the last equation is the only distinction
from the usual spin commutation relations. Accordingly, the
Casimir operator $\hat{\cal S}^{2}$ that commutes with
$\hat{\cal S}_\pm$ and $\hat{\cal S}_z$ should be
defined as
$\hat{\cal S}^{2}\equiv -\hat{\cal S}_+\hat{\cal S}_-
+\hat{\cal S}_z^2-\hat{\cal S}_z$.
The requirement that $\hat{\cal S}_z^2-\hat{\cal S}_z
-\hat{\cal S}^{2}$ must be positive semidefinite leads to
the allowed combinations of eigenvalues
$\{{\cal S}({\cal S}-1),{\cal S}_z\}$ for $\hat{\cal S}^{2}$
and $\hat{\cal S}_z$ such that
${\cal S}=(2N_0+5)/4$
$(N_0=0,1,2,\ldots)$ and
${\cal S}_z={\cal S}+N_2$ $(N_2=0,1,2,\ldots)$.
Here we introduced new
quantum numbers $N_2$ and $N_0$, where the operator
$\hat{\cal S}_+$ raises $N_2$ by one and the relation
$N=2N_2+N_0$ holds. We  may thus interpret $N_2$ as the
number of spin-singlet `pairs', and $N_0$ as that of
all the other bosons.
Noting that the second quantized form of $\hat{P}_0$ is
written as $2\hat{\cal S}_+\hat{\cal S}_-/5$, the second-quantized
form of the spin-dependent
part of the Hamiltonian can be written as
\begin{equation}
\hat{H}=
\frac{c_1}{2}:\hat{\bbox{F}}\cdot\hat{\bbox{F}}:
+\frac{2c_2}{5}\hat{\cal S}_+\hat{\cal S}_-
-p\hat{F}_z,
\label{h2}
\end{equation}
where $c_i\equiv \bar{c}_i\int d\bbox{r}|\phi|^4$.

We first discuss
MFT with a fixed total number of bosons,
and define a state in which all bosons are in
the same single particle state as
$(N!)^{-1/2}(\sum_\alpha\zeta_\alpha\hat{a}_\alpha^\dagger)^N|vac\rangle$,
where $\sum_\alpha|\zeta_\alpha|^2=1$.
Noting that
$\langle\hat{a}_{\alpha^\prime}^\dagger\hat{a}_{\beta^\prime}^\dagger
\hat{a}_\beta\hat{a}_\alpha\rangle=N(N-1)
\zeta_{\alpha^\prime}^*\zeta_{\beta^\prime}^*
\zeta_\beta\zeta_\alpha$, the energy $E_M$ of
the state is written as
\begin{equation}
E_M=
N\left[
\frac{c_1(N-1)}{2}\langle\hat{\bbox{f}}\rangle^2
+\frac{2c_2(N-1)}{5}s^2
-p\langle\hat{f}_z\rangle
\right]
\end{equation}
where
$\langle\hat{\bbox{f}}\rangle^2
=\langle\hat{f}_z\rangle^2
+|2(\zeta_2\zeta_1^*+\zeta_{-1}\zeta_{-2}^*)
+\sqrt{6}\zeta_1\zeta_0^*+\zeta_0\zeta_{-1}^*|^2$,
$\langle\hat{f}_z\rangle
=2(|\zeta_2|^2-|\zeta_{-2}|^2)+|\zeta_1|^2-|\zeta_{-1}|^2$,
and $s^2\equiv
|\zeta_0^2/2-\zeta_1\zeta_{-1}+\zeta_2\zeta_{-2}|^2$.
(An MFT that assumes coherent states with
amplitudes $\{\sqrt N\zeta_\alpha\}$ for the ground state is 
obtained by
replacing the terms $c_i(N-1)$ in $E_M$
by $c_iN$.)
The ground state and its magnetization in MFT are obtained by minimizing
$E_M$, and our results are summarized as follows. When $c_2>0$ and $c_1>0$,
the term including $c_2$ vanishes $(s^2=0)$ for the minimized state,
and 
the magnetization increases linearly with the magnetic field as
$F_z\sim [g\mu_B/c_1]B$. Any Zeeman sublevel can take nonzero 
population
in this case. When $c_2<0$ and $20c_1+|c_2|>0$, 
the $c_2$ term contributes to $F_z$, but it only amounts to
replacing $c_1$ in the expression of $F_z$ above
with $c_1+|c_2|/20$.
This case is quite similar to the spin-1 case, and
only $m=\pm 2$ levels are populated. In other regions of the parameters
$c_1$ and $c_2$, the ground state is ferromagnetic.

Exact energy eigenstates and eigenvalues of 
Hamiltonian (\ref{h2}) can be obtained as follows.
Because operators $\hat{{\cal S}}_\pm$ are invariant
under any rotation of the system, they commute with
$\hat{\bbox{F}}^2$ and $\hat{F}_z$.
The energy eigenstates can thus be classified
according to quantum numbers
$N_2$ and
$N_0$,
total spin $F$,
and magnetic quantum number $F_z$.
We thus denote the eigenstates as
$|N_2,N_0,F,F_z,\lambda\rangle$, where
$\lambda=1,2,\ldots,g_{N_0,F}$ is included to label
degenerate states. The energy eigenvalue for this state
is
\begin{equation}
E=\frac{c_1}{2}[F(F+1)-6N]
+\frac{c_2}{10}(N-N_0)(N+N_0+3)
-pF_z,
\label{e2}
\end{equation}
where we used $2N_2+N_0=N$.
The degeneracy $g_{N_0,F}$ can be calculated from
 generating function~\cite{foot}
\begin{equation}
\sum_{N_0,F}g_{N_0,F}x^{N_0}y^F
=\frac{1-xy+x^2y^2}{(1-xy^2)(1+xy)(1-x^3)}.
\end{equation}
The total spin $F$ can take integer values in the range
$0\le F \le 2N_0$ except for some forbidden values. That is,
$F=1,2,5,2N_0-1$ is not allowed when $N_0=3k(k\in \bbox{Z})$,
and  $F=0,1,3,2N_0-1$ is forbidden when $N_0=3k\pm1$.

It is convenient to consider two cases
separately depending on the
sign of the parameter $c_2$ for the minimization of Eq.\ (\ref{e2}).

{\em (a) $c_2>0$} --- the ground state is
$|N_2=0,N_0=N,F,F_z=F,\lambda\rangle$ with $F$
taking the value closest to $p/c_1-1/2$.
The magnetization $F_z=F$ can take integer values in the range
$0\le F \le 2N$ except for the forbidden values
described above.

The spin correlations in these ground states are rather
complicated in comparison with the spin-1 case. This is
because the condition $\langle\hat{\cal S}_+\hat{\cal
S}_-\rangle=0$ implies that the spin correlation
between {\it any} two particles must avoid the
singlet-like correlation. Except for this constraint,
the spin correlation may be reduced to a combination of
two- and three-particle correlations. Let us define the
operator $\hat{A}^{(n)\dagger}_s$ such that it creates
$n$ bosons in the state with total spin $F=s$ and $F_z=s$
when applied to the vacuum. Consider a set of
unnormalized states,
\begin{equation}
|n_{12},n_{22},n_{30},n_{33}\rangle
=\hat{\bar{P}}_0
(\hat{a}^\dagger_2)^{n_{12}}
(\hat{A}^{(2)\dagger}_2)^{n_{22}}
(\hat{A}^{(3)\dagger}_0)^{n_{30}}
(\hat{A}^{(3)\dagger}_3)^{n_{33}}
|vac\rangle
\label{set}
\end{equation}
with $n_{12},n_{22},n_{30}=0,1,2,\ldots$ and
$n_{33}=0,1$. The operator $\hat{\bar{P}}_0$ is the
projection to the kernel of $\hat{\cal S}_-$, which
ensures $\langle\hat{\cal S}_+\hat{\cal
S}_-\rangle=0$ for these states.
It is easy to see that $|n_{12},n_{22},n_{30},n_{33}\rangle$
are energy eigenstates with $N_2=0$,
$N_0=n_{12}+2n_{22}+3n_{30}+3n_{33}$, and
$F=F_z=2n_{12}+2n_{22}+3n_{33}$.
Note that the states belonging to the same eigenvalue are not
necessarily mutually orthogonal. A further analysis~\cite{foot},
however, shows that these states are linearly independent,
and the degeneracy coincides with $g_{N_0,F}$. The set
(\ref{set}) thus forms a complete basis of the subspace
spanned by $\{|N_2=0,N_0,F,F_z=F,\lambda\rangle\}$.
The form in (\ref{set}) provides an intuitive explanation
for the forbidden values of $F$. For example, $F=0$
is possible only
when $N_0$ is a multiple of 3 because the singlet state is
formed only by three particles.

{\it (b)} $c_2<0$ ---
The ground state should satisfy $F_z=F$.
To determine the remaining parameters $\{N_0,F\}$,
we first separate
$E$ into the part that depends on $F$ and
$l\equiv 2N_0-F$, and the part that depends only on $N$, namely,
\begin{equation}
E=\frac{(20c_1+|c_2|)}{40}g(F,l)
-3c_1N
-\frac{|c_2|}{10} N(N+3),
\end{equation}
where
$g(F,l)=F^2+[1+c(5+2l)-p^{\prime}]F
+cl(l+6)$,
$
c\equiv{|c_2|}/(20c_1+|c_2|)$,
and $p^\prime\equiv{40p}/(20c_1+|c_2|)$.

When $20c_1+|c_2|<0$, the ground state is obtained
by maximizing $g(F,l)$. Suppose first that $N$ is even.
Since $g(F,l)$ is a decreasing function of $l$
in this case, the maximum should be $g(0,0)=0$ or
$g(2N,0)=2N(2N+1+5c-p^\prime)$. The ground state is thus
$\{N_0,F\}=\{0,0\}$ if
$40p <5|c_2|-(2N+1)\left|20c_1+|c_2|\right|$, and
$\{N_0,F\}=\{N,2N\}$ otherwise.
When $N$ is odd, $\{N_0,F\}=\{0,0\}$ is not allowed, and we must
compare $g(0,6)$, $g(2,0)$, and $g(2N,0)$. The ground state is
$\{N_0,F\}=\{1,2\}$ if
$40p <5|c_2|-(2N+3)\left|20c_1+|c_2|\right|$, and
$\{N_0,F\}=\{N,2N\}$ otherwise.
These results indicates that in the small parameter region of
$-5|c_2|/2N\lesssim 20c_1+|c_2|<0$,
magnetization of the
ground state jumps from 0 or 2 to 2N.
Such a large discontinuity does not appear in MFT with a linear
Zeeman potential. (However, in the presence of a quadratic Zeeman
potential, such a jump occurs also in MFT~\cite{Stenger}.)

When $20c_1+|c_2|>0$, the ground state is obtained
by minimizing $g(F,l)$. The function $g(x,0)$ for real $x$ is
minimal when $x=x_0\equiv(p^\prime-5c-1)/2$.
Since $l=0$ is
allowed only  when $F=k^\prime\equiv 2N-4k$ with $k$ being
nonnegative integer, it is sufficient to compare the states 
with
$\{N_0,F\}=\{k^\prime/2-2,k^\prime-4\},\{k^\prime/2,k^\prime-3\},
\{k^\prime/2,k^\prime-2\},
\{k^\prime/2+2,k^\prime-1\},
\{k^\prime/2,k^\prime\}$ when $x_0$ is in the region
$[k^\prime-4,k^\prime]$. The ground state is thus
$\{N_0,F\}=\{k^\prime/2,k^\prime\}$ if
\begin{eqnarray}
\max\{-1-c(2k^\prime-1),-10c(k^\prime+4)\}
<p^\prime-2k^\prime
\nonumber \\
<
\min\{2+2c(3k^\prime+19),3+c(2k^\prime+17)\},
\end{eqnarray}
$\{N_0,F\}=\{k^\prime/2+2,k^\prime-1\}$ if

\begin{equation}
-2+6c(k^\prime+7)
<p^\prime-2k^\prime
<
-10c(k^\prime+4),
\end{equation}
$\{N_0,F\}=\{k^\prime/2,k^\prime-2\}$ if
\begin{eqnarray}
\max\{-5+c(2k^\prime+9),-4-2c(k^\prime-2)\}
<p^\prime-2k^\prime
\nonumber \\
<
\min\{-2+6c(k^\prime+7),-1-c(2k^\prime-1)\},
\end{eqnarray}
and $\{N_0,F\}=\{k^\prime/2,k^\prime-3\}$ if
\begin{equation}
-6+2c(3k^\prime+7)
<p^\prime-2k^\prime
<
-4-2c(k^\prime-2).
\end{equation}
These results indicate how
magnetization increases with the applied magnetic field: When
$F_z\lesssim1/8c$, $F_z$ takes all integer values. In the region
$1/8c\lesssim F_z\lesssim1/4c$,
$F_z$ skips the values $F_z=2N-4k-1$. When
$1/4c\lesssim F_z\lesssim 1/c$, $F_z=2N-4k-3$ are further
skipped, and $F_z$ takes every other integer values.
When $1/c\lesssim F_z$, $F_z=2N-4k$ are the only
allowed values, so $F_z$ increases by 4 at a time.

The reduced form of the states mentioned above is also
helpful to illustrate this behavior. The states with
$F=k^\prime-4,k^\prime-3,k^\prime-2,k^\prime-1,k^\prime$
can be written as
$(\hat{A}^{(2)\dagger}_0)^{k+1}
(\hat{a}^\dagger_2)^{k^\prime/2-2}|vac\rangle$,
$(\hat{A}^{(2)\dagger}_0)^{k}
(\hat{a}^\dagger_2)^{k^\prime/2-3}
\hat{A}^{(3)\dagger}_3|vac\rangle$,
$(\hat{A}^{(2)\dagger}_0)^{k}
(\hat{a}^\dagger_2)^{k^\prime/2-2}
\hat{A}^{(2)\dagger}_2|vac\rangle$,
$(\hat{A}^{(2)\dagger}_0)^{k-1}
(\hat{a}^\dagger_2)^{k^\prime/2-3}
\hat{A}^{(2)\dagger}_2\hat{A}^{(3)\dagger}_3|vac\rangle$,
$(\hat{A}^{(2)\dagger}_0)^{k}
(\hat{a}^\dagger_2)^{k^\prime/2}|vac\rangle$,
respectively. As the energy cost required to break a singlet pair
increases,
transitions accompanied by this breakage requires a stronger field
and are eventually suppressed.

As in the spin-1 case, the exact ground state shows nonzero
population in the $m=0,\pm 1$ levels. Since the expressions of
the exact results for these populations
are lengthy, we only show the leading terms under
the condition $1\ll n_{12} \ll N_2$.
Surprisingly, the populations are considerably different for
the types of possible ground states,
namely,
$(\hat{A}^{(2)\dagger}_0)^{N_2}
(\hat{a}^\dagger_2)^{n_{12}}
(\hat{A}^{(2)\dagger}_2)^{n_{22}}
(\hat{A}^{(3)\dagger}_3)^{n_{33}}|vac\rangle$
with $n_{22}=0,1$ and $n_{33}=0,1$.
The results are
$
\langle\hat{a}^\dagger_1\hat{a}_1\rangle
\sim
\langle\hat{a}^\dagger_{-1}\hat{a}_{-1}\rangle
\sim
N_2(1+n_{33})/n_{12}
$
and
$
\langle\hat{a}^\dagger_0\hat{a}_0\rangle
\sim
N_2(1+2n_{22})/n_{12}.
$
These results indicate that the populations in the $m=0,\pm 1$
states
are very sensitive to the combination of the spin
correlations, and a very small difference
in magnetization leads to large changes in the
populations, by a factor of 2 or 3. The origin of this
drastic change is the bosonic enhancement caused by
the term $(\hat{a}_2^\dagger)^2\hat{a}_{-1}^\dagger$ in
$\hat{A}^{(3)\dagger}_3$ and the term
$\hat{a}_2^\dagger\hat{a}_0^\dagger$ in
$\hat{A}^{(2)\dagger}_2$.

To summarize, we examined magnetic response of
spin-1 and spin-2 BECs by deriving exact eigenstates
of each Hamiltonian with spin-dependent interaction.
The response is stepwise and the spin-1 BEC shows the step
of 2 units reflecting formation or
destruction of singlet-like 
pairs. In the spin-2 case, the spin correlations
among 3 particles appear,
leading to various
step sizes ranging from 1 to 4 units. In a small parameter
region, magnetization jumps from almost zero
to the maximum of the order of $N$.
Some Zeeman-level populations, which are predicted to be zero
in MFT, are found to be nonzero when the magnetic field is
small. These populations decrease rapidly with
the increasing magnetic field,
which can be understood as a consequence of bosonic
enhancement. The bosonic enhancement also serves as an `amplifier'
of a small 
change in spin correlations because it leads to large oscillations
of Zeeman-level populations in the spin-2 BEC.

This work was supported by the Core Research for
Evolutional  Science and
Technology (CREST) of the Japan Science and Technology Corporation
(JST).


\begin{thebibliography}{99}
\bibitem{foot} Details of the derivations will be published elsewhere.
\end{thebibliography}
\end{document}